\documentclass{article}
\usepackage{dcase2017}
\usepackage{amsmath}
\usepackage{graphicx}
\usepackage{times}
\usepackage{booktabs}
\usepackage{subcaption}

\usepackage[hidelinks]{hyperref}
\usepackage{gensymb}
\usepackage{enumitem}
\usepackage{tabularx}
\usepackage{url}

\newcommand{\x}{$\times$}

\title{Multi-Temporal Resolution Convolutional Neural Networks for Acoustic Scene Classification}

\twoauthors
  {Alexander Schindler}
    {
      Austrian Institute of Technology \\
      Center for Digital Safety and Security \\
      Vienna, Austria \\
      alexander.schindler@ait.ac.at}
  {Thomas Lidy, Andreas Rauber}
    {
     Vienna University of Technology\\
     Institute of Software Technology \\
	 Vienna, Austria \\
     {lidy,rauber}@ifs.tuwien.ac.at}

\begin{document}

\ninept
\maketitle

\begin{sloppy}

\begin{abstract}
In this paper we present a Deep Neural Network architecture for the task of acoustic scene classification which harnesses information from increasing temporal resolutions of Mel-Spectrogram segments. This architecture is composed of separated parallel Convolutional Neural Networks which learn spectral and temporal representations for each input resolution. The resolutions are chosen to cover fine-grained characteristics of a scene's spectral texture as well as its distribution of acoustic events. The proposed model shows a 3.56\% absolute improvement of the best performing single resolution model and 12.49\% of the DCASE 2017 Acoustic Scenes Classification task baseline \cite{Mesaros2016_EUSIPCO}.
\end{abstract}


\section{Introduction}
\label{sec:intro}

Convolutional Neural Networks (CNN) \cite{lecun1995convolutional} have become a popular choice in computer vision due to their ability to capture nonlinear spatial relationships which is in favor of tasks such as visual object recognition \cite{krizhevsky2012imagenet}. Their success has fueled interest also in audio-based tasks such as speech recognition and music information retrieval. An interesting sub-task in the audio domain is the detection and classification of acoustic sound events and scenes, such as the recognition of urban city sounds, vehicles, or life forms, such as birds \cite{fazekas2017multi}. The IEEE AASP Challenge DCASE is a benchmarking challenge for the ``Detection and Classification of Acoustic Scenes and Events''. Acoustic Scene Classification (ASC) in urban environments (task 1) is one of four tasks of the 2016 and 2017 competition. The goal of this task is to classify test recordings into one of predefined classes that characterizes the environment in which it was recorded, for example ``metro station'', ``beach'', ``bus'', etc. \cite{Mesaros2016_EUSIPCO}. 

The presented approach attempts to circumvent various limitations of Convolutional Neural Networks (CNN) concerning audio classification tasks. The tasks performed by a CNN are more related to the visual computing domain. A common approach is to use Short-Term Fourier Transform (STFT) to retrieve a Spectrogram representation which is in the following interpreted as a gray-scale image. Commonly a Mel-Transform is applied to scale the Spectrogram to a desired input size. In previous work we have introduced a CNN architecture to learn timbral and temporal representations at once. This architecture takes a Mel-Spectrogram as input and reduces this information in two parallel CNN stacks towards the spectral and the temporal dimension. The combined representations are input to a fully connected layer to learn the concept relevant dependencies. The challenge is how to choose the length of the input analysis window. Acoustic events can be single sounds or compositions of multiple sounds. Acoustic scenes could be described by the presence of a single significant acoustic event such as ship horns for harbors or by combinations of different events. The temporal pattern of such combinations varies distinctively across and within the acoustic scenes (see Figure \ref{fig:example_spectrograms} for examples of acoustic scenes). Choosing the wrong size of the analysis window can either prevent from having sufficient timbral resolution or to fail to recognize acoustic events with longer patterns.

Thus, we propose an architecture that trains on multiple temporal resolutions to harness relationships between spectral sound characteristics of an acoustic scene, and its patterns of acoustic events. This would facilitate to learn more precise representations on a high temporal scale to discriminate timbral differences such as diesel engines from trucks and petrol based engines from private cars. On the other hand, low level temporal resolutions with ranges from several seconds can optimize on different patterns of acoustic events such a speech, steps or passing cars. Finally, the representations of the different temporal resolutions, learned by the parallel CNN stacks, are combined to form an input for a fully connected layer which learns the relationships between them to predict the acoustic scenes annotated in the dataset.


In Section \ref{sec:related} we will give a brief overview of related work. In Section \ref{sec:method} and \ref{sec:augmentation} our method and the applied data augmentation methods are described in detail. Section \ref{sec:eval} describes the evaluation setup and results while results are presented and discussed in Section \ref{sec:results}. Finally, Section \ref{sec:conclusions} summarizes the paper and provides conclusions.

\begin{figure*}[t]
	\centering
	\includegraphics[width=1.00\textwidth]{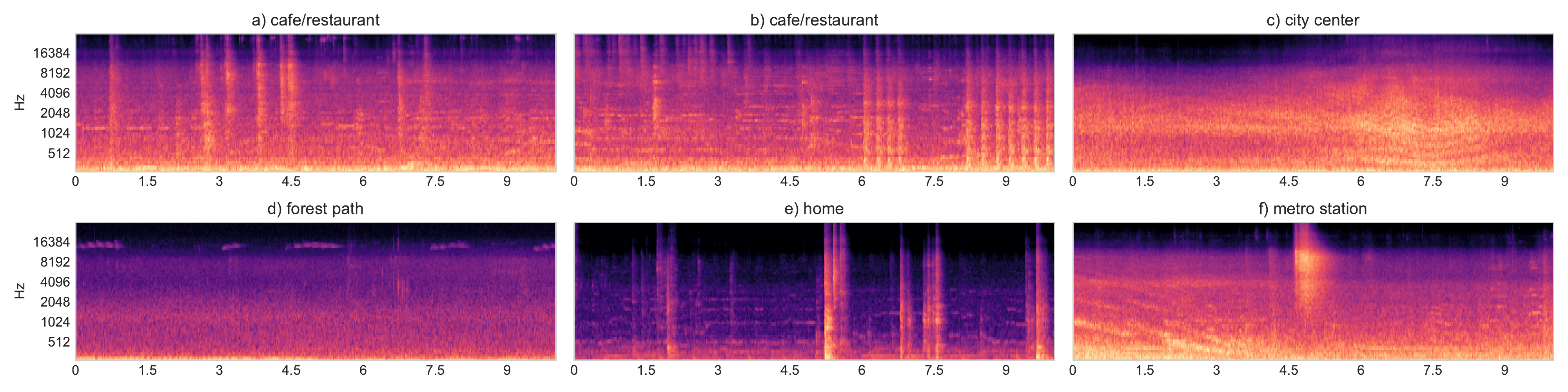}
	\caption{Example Mel-Spectrograms to visualize variances in length and shape of different acoustic events. a) dropping coins into the cash-box, b) beating coffee grounds out of the strainer, c) Doppler effect with Lloyd's mirror effect \cite{lo2002aircraft} of a passing car, d) chirping bird, e) opening and closing of cupboards and drawers in the kitchen, f) arriving subway with pneumatic exhaust.}
	\label{fig:example_spectrograms}
\end{figure*}

\section{Related Work}
\label{sec:related}

The presented approach is based on our DCASE 2016 contribution \cite{Lidy2016} and the modified deeper parallel architecture presented in \cite{schindler2016comparing}. 
Approaches to apply CNNs and Neural Network (NN) architectures to audio analysis tasks were evaluated in \cite{Costa201728}. The authors conclude that DNNs are not yet outperforming crafted feature-based approaches and that best performing results can be achieved through hybrid combinations.
Also the leading contributions to the DCASE 2016 ASC task were not based on DNNs \cite{Eghbal-Zadeh2016,Bisot2016,Park2016}
A similar data augmentation method of mixing audio files of the same class to generate new instances was applied in \cite{takahashi2016deep} and similar perturbations and noise induction was reported in \cite{schluter2015exploring}.
Approaches to ASC using CNN based models were reported in \cite{Santoso2016,Valenti2016}.
Combinations of CNNs with Recurrent Neural Networks (RNN) \cite{xu2017convolutional} have also shown promising results.

\pagebreak

\section{Method}
\label{sec:method}

The presented approach analyses multiple temporal resolutions simultaneously. The design of this architecture is based on the hypothesis that acoustic scenes are composed of the spectral texture or timbre of a scene such as the low-frequent humming of refrigeration units in supermarkets as well as a sequence of acoustic events. These events can be unique for certain acoustic scenes such as the sound of breaking waves at the beach, but usually the characteristics of a scene is described by mixtures of multiple events or sounds. 
Spectral texture or timbre analysis requires high temporal resolutions. To distinguish the trembling fluctuations of a truck's diesel engine from a private car an analysis window of several milliseconds is required. Acoustic events, as exemplified in Figure \ref{fig:example_spectrograms}, happen on a much broader temporal scale. The pattern of beating the coffee grounds out of the strainer of an espresso machine in a caffee (see Figure \ref{fig:example_spectrograms} b) requires an analysis window of 0.5 to 1 seconds. Up to 5 seconds are required for the very significant dropping sound of a decelerating Metro engine with the pneumatic exhaust of the breaks at full halt (see Figure \ref{fig:example_spectrograms} f).

Figure \ref{fig:window_size_example} visualizes different spectral resolutions at a fixed start-offset from audio content recorded in a residential area. Figure \ref{fig:window_size_example} a) visualizes the low-frequent urban background hum at a very high temporal resolution. At this level a CNN can learn a good timbre representation for acoustic scenes, but it is not able to recognize acoustic events that are longer than 476 milliseconds. Patterns such as speech (see Figure \ref{fig:window_size_example} c) or combinations of patterns such as people talking while a car is passing (see Figure \ref{fig:window_size_example} e) require much longer analysis windows, up to several seconds.
The problem with single-resolution CNNs is, that a decision has to be made concerning the length and precision of the analysis window. A high temporal resolution prevents from recognizing long events while a low resolution is not able to effectively describe timbre. Increasing the size of the input segment to widen the analysis window would also increase the size of the model, its number of trainable parameters and the number of required training instances to avoid over-fitting. If pooling-layers are extensively used to reduce the size of the model, the advantage of the high temporal resolution gets lost in these data-reduction steps.

Thus, we propose to use multiple inputs at different temporal resolutions to have separate CNN models learn acoustic scene representations at different scales which are finally combined to learn the categorical concepts of the acoustic scene classification dataset.

\begin{figure*}[t] 
\centering
\includegraphics[width=1.00\textwidth]{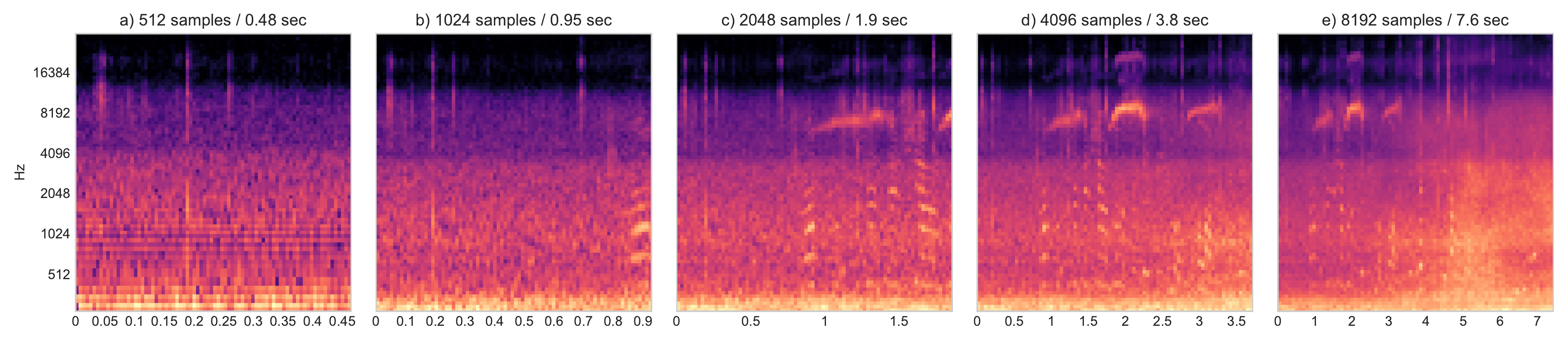}
\caption{Input Segments for the Convolutional Neural Networks with 80 Mels spectral and five different temporal resolutions with fixed start-offset. a) spectral texture of residential area background noise, b) person saying a word (vertical wave-line), c) person talking, tweet of a bird (horizontal arc), d) person talking, bird tweeting, e) person talking, bird tweeting, car passing (light cloud to the right)}
\label{fig:window_size_example}
\end{figure*}

\subsection{Deep Neural Network Architecture}

The presented architecture consists of identical but not shared Convolutional Neural Network (CNN) stacks - one for each temporal resolution. These stacks are based on the parallel architectures initially described in \cite{pons_cbmi2016} and further developed in \cite{Lidy2016,schindler2016comparing,fazekas2017multi}. The fully connected output layers of each parallel CNN stack, which is considered to contain the learned representation for the corresponding temporal resolution, are combined to the multi-resolution model.

\begin{description}[leftmargin=.4cm,itemsep=0pt,labelwidth=.4cm,labelindent=0cm,rightmargin=0.1cm] 

\item[The Parallel Architecture:]
This architecture uses a parallel arrangement of CNN layers with rectangular shaped filters and Max-Pooling windows to capture spectral and temporal relationships at once \cite{pons_cbmi2016}. The parallel CNN stacks use the same input - a log-amplitude transformed Mel-Spectrogram with 80 Mel-bands spectral and 80 STFT frames temporal resolution. The variant used in this paper (see Figure \ref{fig:parallel_cnn}) is based on the deep architecture presented in \cite{schindler2016comparing}. The first level of the parallel layers are similar to the original approach \cite{Lidy2016}. They use filter kernel sizes of 10\x 23 and 21\x 10 to capture frequency and temporal relationships. To retain these characteristics the sizes of the convolutional filter kernels as well as the feature maps are sub-sequentially divided in halves by the second and third layers. The filter and Max Pooling sizes of the fourth layer are slightly adapted to have the same rectangular shapes with one part being rotated by 90\degree. Thus, each parallel layer subsequently reduces the input shape to 2 \x 10 dimensions - one layer reduces the spectral while preserving the temporal information, the other performs the same reduction on the temporal axis. The final equal dimensions of the final feature maps of the parallel model paths balances their influences on the following fully connected layer with 200 units.


\item[Multi-Temporal Resolutions CNN:]
The proposed architecture instantiates one parallel architecture for each temporal resolution (see Figure \ref{fig:parallel_cnn}. Their fully connected output layers are concatenated. To learn the dependencies between the sequences of spectral and temporal representations of the different temporal resolutions an intermediate fully connected layer with 512 units is added before the Softmax output layer.

\item[Dropping out Resolution Layers:]
To support the final fully connected layer in learning relations between the different resolutions, a layer that has been added that drops out entire resolutions of the concatenated intermediate layer of the multi-resolution architecture.

\end{description}

\section{Data Augmentation}
\label{sec:augmentation}

The most challenging characteristics of the provided dataset is its low variance. Table \ref{tab:dataset_overview} depicts that for each class audio content of 3120 seconds length is provided. Nevertheless, this content originates from only 13 to 18 different locations per class. To create more data instances these recordings have been split into 10 seconds long audio files, but this does not introduce more variance due to very high self-similarity within a location. This low variance leads complex neural networks with a large number of trainable parameters to over-fit on the training data. Further, the limitation of 10 seconds per file prevents from using larger analysis windows. To circumvent these shortcomings data augmentation using the following methods is applied:

\begin{description}[leftmargin=.4cm,itemsep=0pt,labelwidth=.2cm,labelindent=0cm,rightmargin=0.1cm] 

\item[Split-Shuffle-Remix of audio files:] 
To create additional audio content by increasing the length of an audio file its content is segmented by non-silent intervals. To create approximately 10 segments the Decibel-threshold is iteratively increased until the desired quantity is reached. These segments are duplicated to retrieve four identical copies which corresponds to 40 seconds of audio. All segments are then randomly reordered and remixed into a final combined audio file.

\item[Remixing Places:] 
To introduce more variance in the provided data, additional training examples are created by mixing files of the same class. Based on the assumption that classes are composed of a certain spectral texture and a set of acoustic events, mixing files of the same class would generate new recordings of this class. For each possible pairwise combination of locations within a class, a random file for each location is selected. The recordings are mixed by averaging both signals.

\item[Pitch-Shifting:] 
The pitch of the audio signal is increased or decreased within a range of 10\% of its original frequency while keeping its tempo the same. The 10\% range has been subjectively assessed. Larger perturbations sounded unnaturally.

\item[Time-Stretching:] 
The audio signal is speed up or slowed down randomly within a range of 10\% at maximum of the original tempo while keeping its pitch unchanged.

\item[Noise Layers:] 
A data-independent augmentation method to increase the model's robustness. The input data is corrupted by adding Gaussian noise with a probability of $\sigma = 0.1$ is to the Mel-Spectrograms. The probability $\sigma$ has been empirically evaluated in preceding experiments using different single-resolutions models. From the tested values [0.05, 0.1, 0.2, 0.3] a $\sigma$ of 0.1 improved the model's accuracies most.

\end{description}

\begin{figure}[t]
\centering

\begin{subfigure}[b]{0.7\columnwidth}
   \includegraphics[width=\columnwidth]{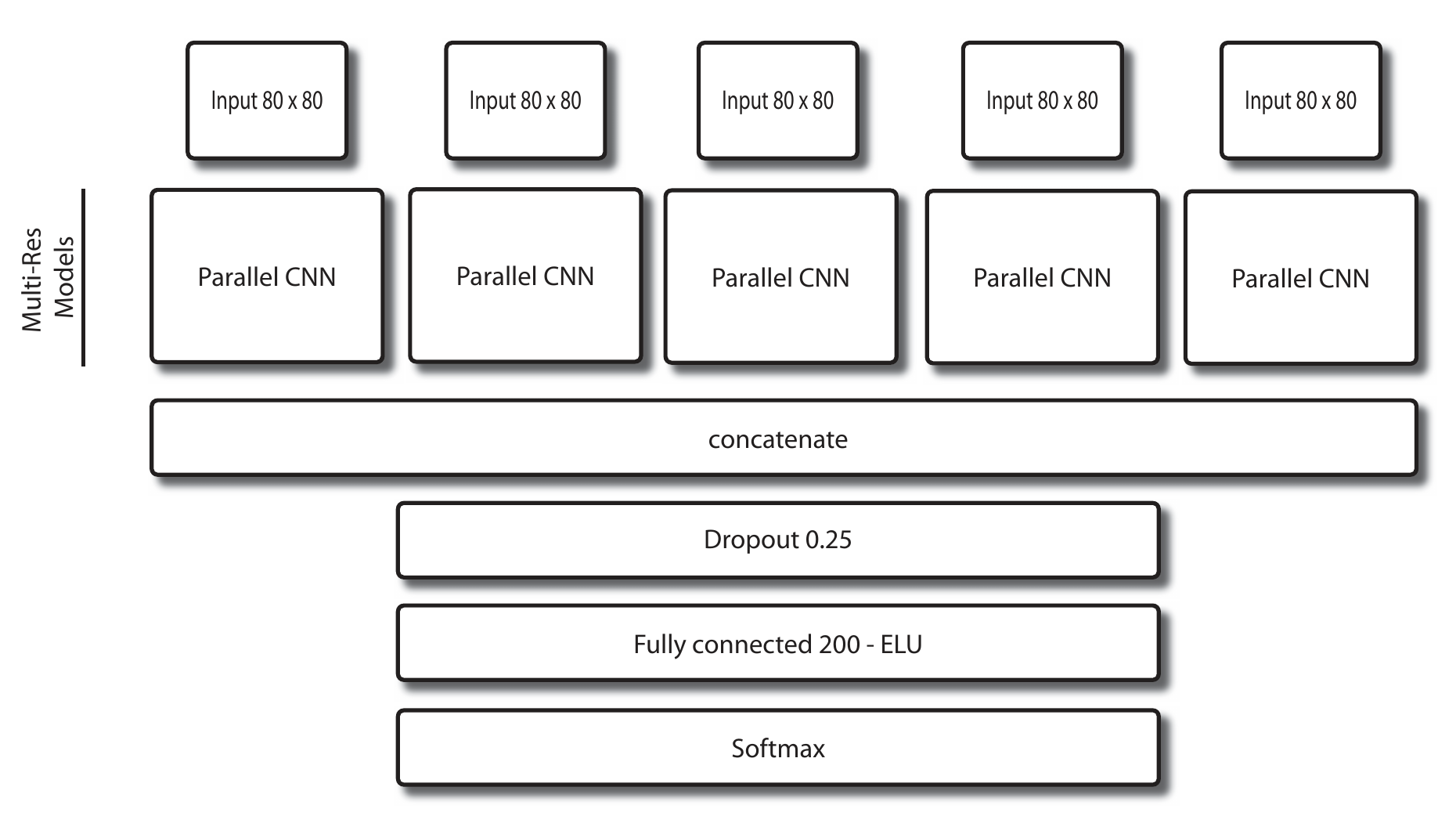}
   \caption{Multi-Resolution Model}
   \label{fig:mode}
\end{subfigure}

\begin{subfigure}[b]{0.6\columnwidth}
   \includegraphics[width=\columnwidth]{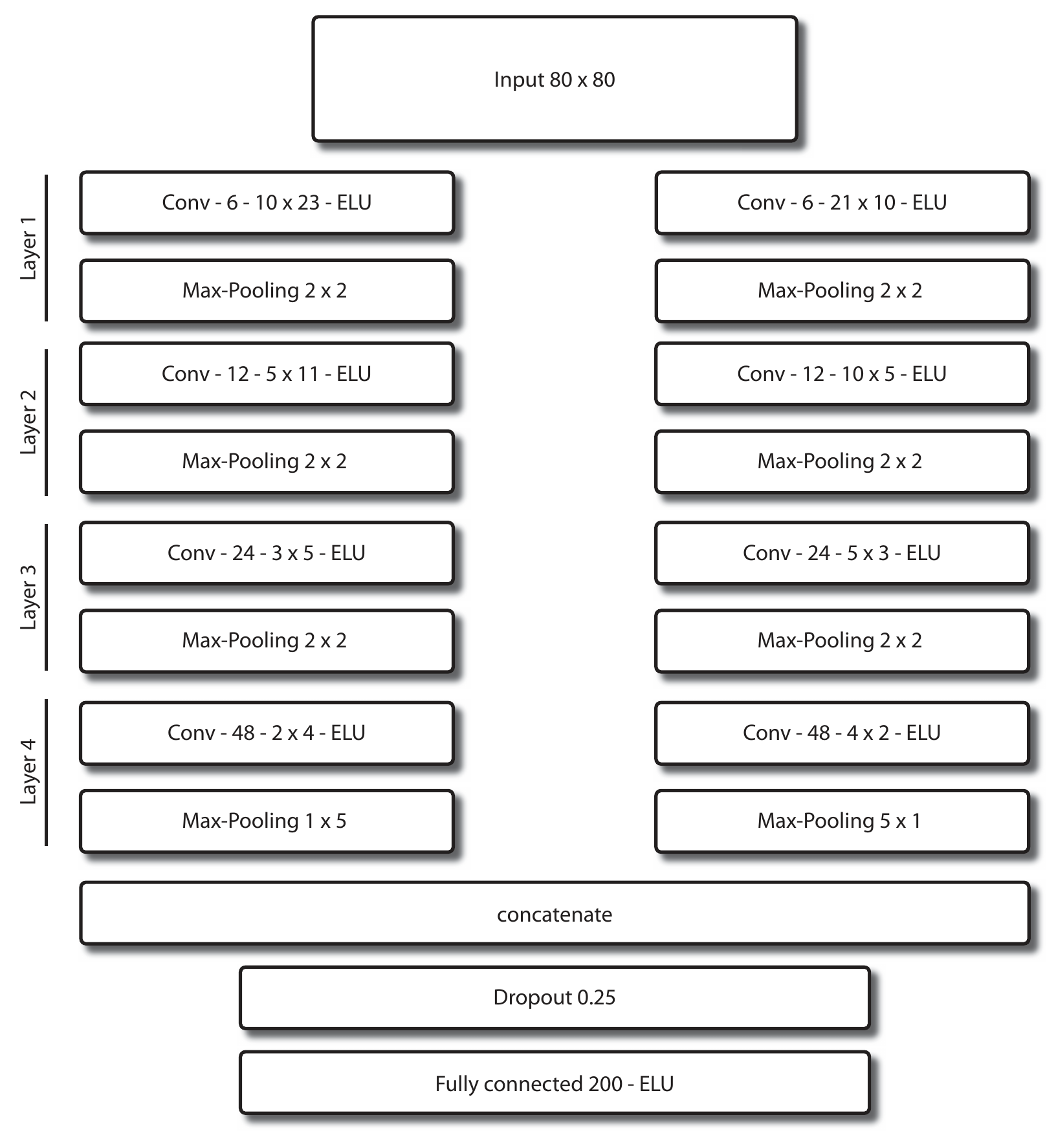}
   \caption{Parallel CNN Architecture}
   \label{fig:parallel_cnn} 
\end{subfigure}

\caption[Two numerical solutions]{The \textit{Multi-Resolution Model} (a) which consists of one \textit{Parallel CNN Architecture} (b) per temporal resolution.}

\vskip -2mm
\end{figure}

\section{Evaluation}
\label{sec:eval}

The presented approach was evaluated on the \textit{development dataset} of the \textit{TUT Acoustic Scenes 2017 dataset} \cite{Mesaros2016_EUSIPCO}. The dataset consists of 15 classes representing typical urban and rural acoustic scenes (see Table \ref{tab:dataset_overview}).
%
%
4-fold cross-validation was applied using grouped stratification which preserved the class distribution of the original ground-truth assignment in the train/test splits as well as ensured that files of the same location are not split across them.
%
%
The performance was measured in classification accuracy on a per-instance-level (\textit{raw}) for every extracted Mel-Spectrogram as well as on a per-file-level (\textit{grouped}) by calculating the average Softmax response for all Mel-Spectrograms of a file.
%
\noindent
For each audio file 10 log-amplitude scaled Mel-Spectrograms with 80 Mels times 80 frames are extracted from the normalized input signal using random offsets and increasing FFT window sizes of 512, 1024, 2048, 4096, 8192 samples with 50\% overlap. To augment the data, additional 10 random input segments were extracted for \textit{time-stretched}, \textit{pitch-shifted} \textit{place-wise remixed} audio content. \textit{Split-Shuffle-Remix} augmentation preceded all feature extraction processes. 
The neural networks were trained using Nadam optimization \cite{dozat2016incorporating} with \textit{categorical crossentropy loss} at $10^{-5}$ learning rate and a batch-size of 32. The learning rate was reduced by 10\% if the validation loss did not improve over 3 epochs maintaining a minimum rate of $5*10^{-6}$.

The evaluation is divided into single- and multi-resolution experiments. First, for each of the combined model's resolutions a separate \textit{parallel CNN model}, and second, the full multi-resolution model is evaluated. Both experiments are performed using un-augmented (\textit{raw}) and augmented input data.

\begin{table}[]
\centering
\caption{Per class dataset Overview. Number of different locations, complete length as well as min/max/mean length of audio content.}
\label{tab:dataset_overview}
\scriptsize
\begin{tabularx}{0.9\columnwidth}{l|c|rrrr}

\toprule
Label &  num diff  & \multicolumn{4}{c}{Audio length (in seconds)} \\
      &  locations &  sum & min &  max &  mean \\

\midrule
beach             &             17 &         3120 &          120 &          210 &         183.5 \\
bus               &             18 &         3120 &           60 &          300 &         173.3 \\
cafe/restaurant   &             16 &         3120 &          120 &          300 &         195.0 \\
car               &             17 &         3120 &           90 &          270 &         183.5 \\
city\_center      &             15 &         3120 &          150 &          270 &         208.0 \\
forest\_path      &             18 &         3120 &           60 &          300 &         173.3 \\
grocery\_store    &             17 &         3120 &          120 &          270 &         183.5 \\
home              &             16 &         3120 &           90 &          300 &         195.0 \\
library           &             16 &         3120 &          150 &          240 &         195.0 \\
metro\_station    &             17 &         3120 &           90 &          300 &         183.5 \\
office            &             13 &         3120 &          150 &          300 &         240.0 \\
park              &             17 &         3120 &          120 &          210 &         183.5 \\
residential\_area &             17 &         3120 &          120 &          240 &         183.5 \\
train             &             17 &         3120 &           90 &          270 &         183.5 \\
tram              &             17 &         3120 &           60 &          300 &         183.5 \\
\bottomrule
\end{tabularx}
\vskip -6mm
\end{table}

\section{Results and Discussion}
\label{sec:results}

As shown in Table \ref{tab:results} and Figure \ref{fig:class_acuracies} the proposed multi-resolution model clearly outperforms the best performing single-resolution models by 3.56\%. Especially the classes \textit{train, metro\_station, residential\_area} and \textit{cafe/resaturant} indicate that the model harnesses dependencies between the temporal resolutions. Although an improvement can already be observed on un-augmented (\textit{raw}) data, the high complexity of the model especially gains from the added variance of augmented data. An interesting observation though is that the augmentation had no or a slightly degrading effect on the classes \textit{car, grocery\_store} and \textit{city\_center}, which seem to be unaffected or distorted by timbral and temporal perturbations or by mutual remixing.
Grouping and averaging the predictions for a file of all single-resolution models (see \textit{'grouped single'} in Table \ref{tab:results}) does not increase the performance of these models, nor is it comparable to the multi-resolution model.
It was further observed that lower temporal resolutions perform better than higher. This could indicate that the higher contrast of peaking spikes in the spectrograms makes it easier for algorithms to learn better and more discriminative representations than from the noise-like pattern of higher temporal resolutions.
As already reported in preceding studies \cite{Lidy2016,Lidy_Schindler_MIREX2016,schindler2016comparing} the grouped accuracy outperforms instance based (\textit{raw}) prediction. Averaging over multiple predicted segments of a test file balances outliers in the classification results.
The custom dropout which dropped the output of two random resolution CNN stacks showed little effect on the general performance of a model. Conventional Dropout with a probability of $\sigma = 0.25$ seemed sufficient.

\begin{table}[t]
\centering
\scriptsize
\caption{Experimental results (classification accuracy with standard deviation over cross-validation folds). Single-resolution model results provided on top, multi-resolution models at the bottom.}
\label{tab:results}
\begin{tabularx}{0.9\columnwidth}{lllll}
\toprule
fft            & instance      &  grouped      & instance      &   grouped     \\
win size       & raw           &  raw          & augmented     &   augmented   \\
\midrule
512            &  64.14 (2.84) &  70.32 (2.96) &  69.06 (4.33) &  76.63 (4.44) \\
1024           &  66.32 (2.58) &  71.27 (3.06) &  71.70 (5.46) &  77.06 (5.46) \\
2048           &  66.83 (1.52) &  70.23 (1.99) &  76.24 (2.53) &  80.46 (3.30) \\
4096           &  69.50 (2.83) &  71.92 (3.23) &  79.20 (3.03) &  81.66 (3.29) \\
8192           &  69.66 (2.58) &  71.47 (2.95) &  82.26 (2.40) &  83.73 (2.63) \\
\hline
grouped single &               &  73.12        &               &  83.19        \\
\hline
multi-res      &  72.23 (4.15) &  74.30 (4.81) &  85.22 (2.11) &  87.29 (2.02) \\
multi-res do   &  69.39 (2.77) &  72.05 (3.26) &  82.51 (2.37) &  86.04 (3.03) \\
\bottomrule
\end{tabularx}
\end{table}

\begin{figure}[t!]
	\centering
	\includegraphics[width=1.0\columnwidth]{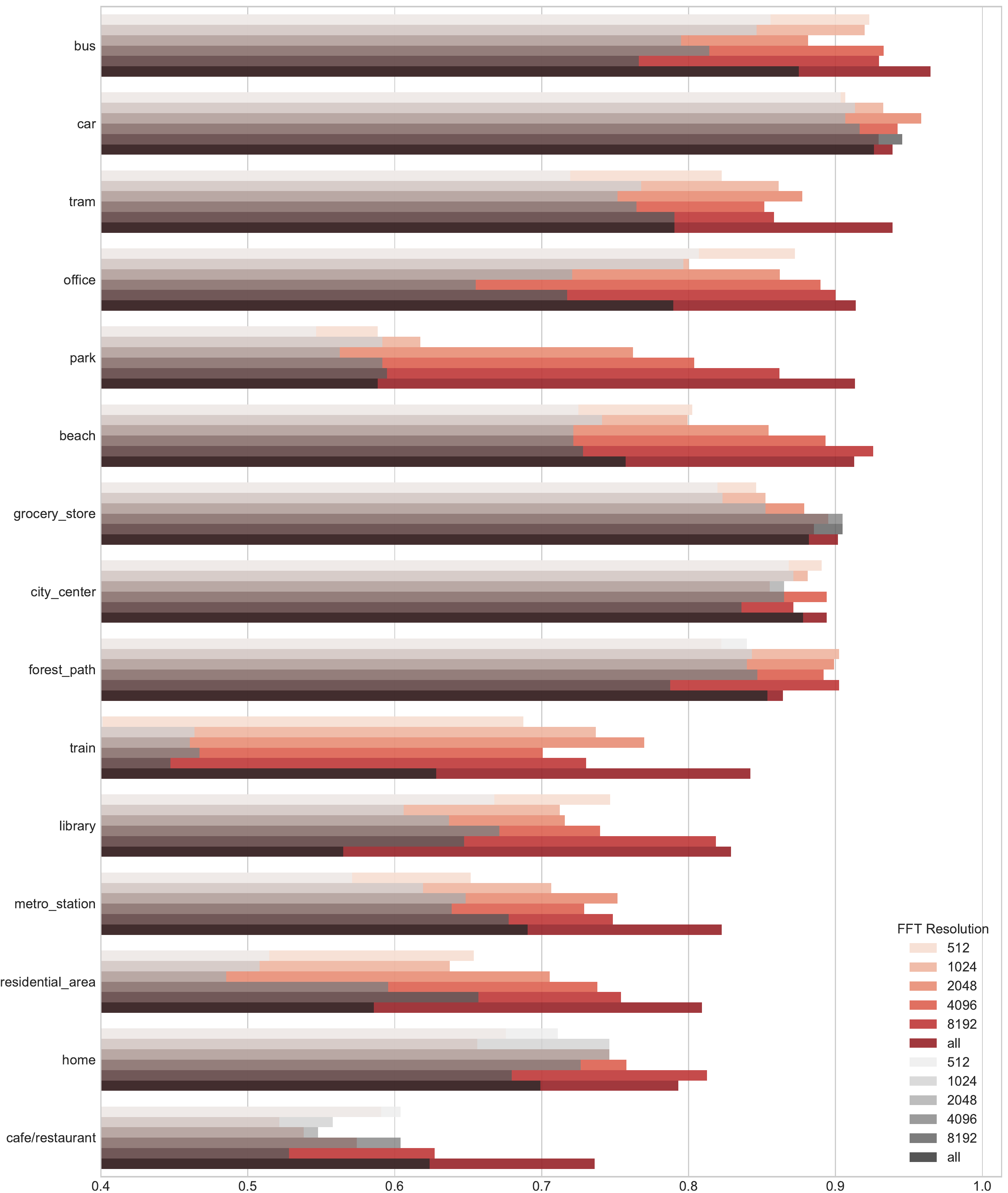}
	\caption{Results per class and FFT window size with ascending temporal resolutions. Multi-resolution results at last. Grayed bars represent un-augmented data, red bars augmented.}
	\label{fig:class_acuracies}
\vskip -6mm
\end{figure}

\section{Conclusions and Future Work}
\label{sec:conclusions}

The presented study introduced a Convolutional Neural Network (CNN) architecture which harnesses multiple temporal resolutions to learn dependencies between timbral properties of an acoustic scene as well as its temporal pattern of acoustic events. The experimental results showed that the proposed multi-resolution model outperforms the all single-resolution and combined models by at least 3.56\%. Future work woul concentrate on improved data augmentation models, including evaluations on which augmentation methods have an improving/degrading effect on the classes (e.g. grocery store) and which methods can be applied to make the lower performing classes more discriminative.

\bibliographystyle{IEEEtran}
\bibliography{refs}

\end{sloppy}
\end{document}